\documentclass[final,5p,times,twocolumn]{elsarticle}

\usepackage{amssymb}
\usepackage{amsmath}

\usepackage{hyperref}
\usepackage[capitalise]{cleveref}
\crefname{appendix}{}{} \usepackage{booktabs}
\usepackage{siunitx}
\usepackage{bm} 

\usepackage{tikz}
\IfFileExists{.externalize}{\tikzexternalize[prefix=tikz_external/]
  \tikzexternalize
}{}
\usepackage{pgfplots}

\newcommand{\pder}[2]{\ensuremath{\frac{\partial #1}{\partial #2}}}

\journal{}

\newcommand{\ppmat}[1]{\ensuremath{\mathbf{#1}}}
\newcommand{\ppvec}[1]{\ensuremath{\bm{#1}}}
\newcommand{\filter}[1]{\ensuremath{\overline{#1}}}
\newcommand{\avg}[1]{\ensuremath{\langle{#1\rangle}}}
\newcommand{\graph}{\ensuremath{\mathcal{G}}}
\newcommand{\edges}{\ensuremath{\mathcal{E}}}
\newcommand{\vertices}{\ensuremath{\mathcal{N}}}
\newcommand{\model}{\ensuremath{\mathcal{M}}}
\newcommand{\cs}{\ensuremath{C_{\!s}}}
\newcommand{\cw}{\ensuremath{C_{\!w}}}
\newcommand{\dtrl}{\ensuremath{\Delta t_{\text{RL}}}}
\newcommand{\tend}{\ensuremath{t_{\text{end}}}}
\newcommand{\tdns}{\ensuremath{t_{\text{DNS}}}}
\newcommand{\nepochs}{\ensuremath{n_{\text{epochs}}}}
\newcommand{\nenvs}{\ensuremath{n_{\text{envs}}}}
\newcommand{\kmax}{\ensuremath{k_{\text{max}}}}
\newcommand{\Edns}{\ensuremath{\filter{E}_{\text{DNS}}}}
\newcommand{\Eles}{\ensuremath{E_{\text{LES}}}}
\newcommand{\Udns}{\ensuremath{\avg{U_{\text{DNS}}}}}
\newcommand{\Ules}{\ensuremath{\avg{U_{\text{LES}}}}}
\newcommand{\Unew}{\ensuremath{\avg{U_{\text{new}}}}}
\newcommand{\Sij}{\ensuremath{S_{\! ij}}}
\newcommand{\commerror}{\ensuremath{\mathcal{C}}}
\newcommand{\discop}{\ensuremath{\mathcal{R}}}
\newcommand{\rewardsmooth}{\ensuremath{\beta}}
\newcommand{\state}{\ensuremath{s}}
\newcommand{\states}{\ensuremath{\mathcal{S}}}
\newcommand{\action}{\ensuremath{a}}
\newcommand{\actions}{\ensuremath{\mathcal{A}}}

\newcommand{\reward}{\ensuremath{r}}
 
\begin{document}

\begin{frontmatter}

\title{Harnessing Equivariance: Modeling Turbulence with Graph Neural Networks}

\author[cwi]{Marius Kurz\corref{cor1}} \ead{marius.kurz@cwi.nl}
\author[iag]{Andrea Beck} \ead{andrea.beck@iag.uni-stuttgart.de}
\author[cwi,eindhoven]{Benjamin Sanderse} \ead{benjamin.sanderse@cwi.nl}

\affiliation[cwi]{organization={Centrum Wiskunde \& Informatica},
          addressline={Science Park 123},
          city={Amsterdam},
          postcode={1098 XG},
country={The Netherlands}
          }
\affiliation[iag]{organization={Institute of Aerodynamics and Gas Dynamics, University of Stuttgart},
          addressline={Wankelstraße 3},
          city={Stuttgart},
          postcode={70563},
country={Germany}
          }
\affiliation[eindhoven]{
    organization={Centre for Analysis, Scientific Computing and Applications, Eindhoven University of Technology},
    addressline={PO Box 513}, 
    city={Eindhoven},
    postcode={5600 MB}, 
country={The Netherlands}
}
\cortext[cor1]{Corresponding author}

\begin{abstract}
This work proposes a novel methodology for turbulence modeling in Large Eddy Simulation~(LES) based on Graph Neural Networks~(GNNs), which embeds the discrete rotational, reflectional and translational symmetries of the Navier-Stokes equations into the model architecture.
  In addition, suitable invariant input and output spaces are derived that allow the GNN models to be embedded seamlessly into the LES framework to obtain a symmetry-preserving simulation setup.
The suitability of the proposed approach is investigated for two canonical test cases: Homogeneous Isotropic Turbulence~(HIT) and turbulent channel flow.
  For both cases, GNN models are trained successfully in actual simulations using Reinforcement Learning~(RL) to ensure that the models are consistent with the underlying LES formulation and discretization.
  It is demonstrated for the HIT case that the resulting GNN-based LES scheme recovers rotational and reflectional equivariance up to machine precision in actual simulations.
  At the same time, the stability and accuracy remain on par with non-symmetry-preserving machine learning models that fail to obey these properties.
  The same modeling strategy translates well to turbulent channel flow, where the GNN model successfully learns the more complex flow physics and is able to recover the turbulent statistics and Reynolds stresses.
  It is shown that the GNN model learns a zonal modeling strategy with distinct behaviors in the near-wall and outer regions.
  The proposed approach thus demonstrates the potential of GNNs for turbulence modeling, especially in the context of LES and RL.
\end{abstract}


\end{frontmatter}

\section{Introduction}\label{sec:introduction}
The last decade has seen a massive surge in the application of machine learning (ML) to various tasks in computational fluid dynamics (CFD)~\cite{brunton2020machine,vinuesa2022enhancing}.
More recently, increasing attention has been paid to augment the standard ``black-box'' approach of ML by integrating prior knowledge and constraints of the underlying physical processes into the modeling to yield more interpretable and physically meaningful models~\cite{karniadakis2021physicsinformed}.
These constraints include for instance adhering to a set of governing differential equations using the Physics-Informed Neural Networks (PINN) paradigm~\cite{raissi2019physics}, encoding fundamental equivariance and symmetry properties into the model architecture~\cite{ling2016reynolds}, or designing data-driven models such that they satisfy the conservation of energy by construction~\cite{vangastelen2024energyconserving}.
The rationale behind these efforts is that incorporating prior knowledge in the form of physical constraints into ML models has shown to yield preferable properties in practice even though this relation is oftentimes based more on intuition and empirical evidence than on a rigorous theoretical foundation~\cite{sanderse2024scientific}.
For instance, it has been demonstrated that the integration of physical constraints can reduce the amount of training data required~\cite{guan2023learning}, improve the models' accuracy and long-term stability for forecasting~\cite{shankar2023importance,toshev2023e3} or result in better generalization abilities~\cite{wang2021incorporating} in comparison to vanilla ML models that have to learn these physical relationships implicitly from data.

More recently, Graph Neural Networks (GNNs)~\cite{kipf2017semisupervised} have gained popularity in CFD, since their graph-based representation of data fits naturally the mesh-based as well as particle-based methods commonly used in CFD~\cite{sanchezgonzalez2020learning}.
With this, they address major shortcomings of commonly employed fully-connected or convolutional neural network (CNN) architectures that suffer from a fixed input size and a limitation to regular grid structures, respectively \cite{cheng2025machine}.
Besides their ability to handle non-trivial geometries and unstructured grids, GNNs can also be designed to embed the geometrical symmetries of the underlying equations into the model architecture as a hard constraint.
For instance, \citet{toshev2023e3} demonstrated that GNNs can be used to construct E(3)-equivariant particle methods that clearly outperform their non-equivariant counterparts in terms of accuracy and stability. 
\citet{lino2022multiscale} used GNNs to directly predict the solution of the unsteady Euler equations in a multiscale setting, where the GNNs designed to respect the rotational symmetries of the Euler equations showed better accuracy and generalization capabilities than non-equivariant models.

First studies also employed GNNs for turbulence modeling.
Here, it is widely recognized that oftentimes some form of \textit{a posteriori} training with actual simulations is required to ensure that the closure models are consistent with the underlying equations, in particular for implicitly filtered Large Eddy Simulations~(LES)~\cite{sanderse2024scientific}.
This is in contrast to \textit{a priori} training based on snapshots of high-fidelity simulations, which can potentially lead to inconsistencies between the precomputed training data and the actual simulation environment~\cite{kurz2021investigating}.
Some common methods for \textit{a posteriori} training entail data assimilation techniques~\cite{rasp2020coupled}, Reinforcement Learning~(RL)~\cite{beck2023toward}, or differentiable programming~\cite{list2022learned,agdestein2025discretize}.
To this end, \citet{quattromini2025active} combined GNNs with data assimilation to learn the subgrid forces for the two-dimensional RANS equations for the flow around differently shaped bluff bodies.
\citet{kim2024generalizable} used a GNN to predict the subgrid stress tensor in a two-dimensional LES setting.
They implemented custom adjoints for their solver to train the GNN within the actual simulation in an end-to-end differentiable manner.
However, both works employ GNNs mainly for their ability to adhere to non-trivial geometries and not to incorporate the fundamental symmetries of the physics into the model architecture.
In contrast, \citet{dupuy2023modeling, dupuy2024using} used GNNs to train wall models for LES on unstructured grids in a purely supervised manner.
While they addressed the symmetries of the model, its geometric symmetries, i.e.\ rotation and reflection, are only enforced weakly via data augmentation and not by the model architecture itself.
Note that while we focus in this discussion on data-driven or data-augmented modeling, these properties are equally important in classical closure models.
At the same time, it is well known that these existing, classical models also often disobey or disregard these constraints~\cite{oberlack1997invariant}.

We address these limitations of former modeling strategies and devise a training framework for closure modeling in three-dimensional LES that yields accurate and long-term stable simulation results, while recovering rotational, translational and reflectional symmetries up to machine precision.
The contributions of this work are the following.
First, we propose a novel strategy based on GNNs that allows to embed these symmetries into the data-driven model as a hard constraint.
Second, we design the input and output space of the GNNs such that they retain  the symmetries of the model.
Third, this creates a modeling strategy that can be embedded seamlessly into the LES solvers independent of their specific distribution of solution points without having to rely on interpolating the flow field to Cartesian grids, specific domain sizes or the computation of global flow quantities.
In this work, a modern, high-order discontinuous Galerkin (DG) method~\cite{chapelier2024comparison} serves as underlying discretization scheme, but the proposed modeling strategy is not limited to this particular choice.
Fourth, we combine this modeling strategy with RL to train the GNNs in an end-to-end fashion in actual simulations to avoid inconsistencies between artificial training data and real simulations~\cite{kurz2023deep,beck2023toward}.
Fifth, we demonstrate the suitability of the proposed approach for two canonical, three-dimensional test cases: homogeneous isotropic turbulence (HIT) and turbulent channel flow.
Moreover, we demonstrate that the trained GNN models recover the fundamental symmetries of the underlying equations in actual simulations.
To the best of our knowledge, this is the first work addressing three-dimensional LES modeling with GNNs and the first to combine GNNs and their advantages with RL for turbulence modeling.

This work is organized as follows.
First, \cref{sec:preliminaries} introduces the closure problem of turbulence and discusses the importance of the fundamental symmetries of the NSE and LES equations.
Based on this, we outline our novel modeling approach in \cref{sec:equivariance}.
This approach employs GNNs to embed those fundamental symmetries into the data-driven models.
The implementation details of the model architecture and its training using an RL approach are summarized in \cref{sec:implementation}.
The modeling strategy is then applied to two canonical test cases for turbulent flow, HIT and turbulent channel flow, in \cref{sec:hit} and \cref{sec:channel}, respectively.
Finally, \cref{sec:conclusion} summarizes the main findings and provides an outlook on future work.
 
\section{Preliminaries}\label{sec:preliminaries}
Before introducing our novel approach for equivariant LES modeling, we first introduce the closure problem in \cref{sec:closure}.
Based on this, we discuss the symmetries of the governing equations and the importance of incorporating these into data-driven turbulence models in \cref{sec:symmetries}.

\subsection{Stating the Closure Problem}\label{sec:closure}

Turbulence is a chaotic, multi-scale phenomenon characterized by a wide range of spatial and temporal scales.
The evolution of turbulent flows is governed by the compressible Navier-Stokes Equations (NSE), which can be written in their conservative form as
\begin{equation}
  \pder{\ppvec{U}}{t} + \nabla \cdot \ppmat{F}(\ppvec{U}) = 0,
  \label{eq:nse}
\end{equation}
with the conserved variables $\ppvec{U}=(\rho, \rho \ppvec{u},\rho e)^T$ entailing the mass, momentum and energy density, respectively.
The flux matrix $\ppmat{F}(\ppvec{U})$ encompasses the nonlinear convective and linear viscous contributions, where each of its three columns $i=1,2,3$ corresponds to the flux vector in one principal direction written as
\begin{equation}
  \ppvec{F}_i(\ppvec{U}) = \begin{pmatrix}
    \rho u_i \\
    \rho u_1 u_i + \delta_{1i} p - \tau_{1i} \\
    \rho u_2 u_i + \delta_{2i} p - \tau_{2i} \\
    \rho u_3 u_i + \delta_{3i} p - \tau_{3i} \\
    (\rho e + p) u_i - \tau_{ij}u_j -q_j
  \end{pmatrix},
\label{eq:nse_fluxes}
\end{equation}
with the pressure $p$, the viscous stress tensor $\ppmat{\tau}$ and the heat flux vector $\ppvec{q}$, which can be computed from $\ppvec{U}$ using suitable constitutive relations.
Moreover, $\delta_{ij}$ denotes the Kronecker delta.
While turbulence can in principal be predicted accurately by solving \cref{eq:nse} directly, its multi-scale nature imposes prohibitive resolution requirements on Direct Numerical Simulations~(DNS) of most turbulent flows of interest.
One approach to reduce this computational cost is the framework of LES.
The LES methodology alleviates the resolution requirements of turbulent flow by only resolving the large, energy-containing scales of the flow, while modeling the influence of the dynamics of the fine scales.
For this, a spatial coarse-graining filter $\filter{(\cdot)}$ is applied to \cref{eq:nse}, which yields the LES equations as
\begin{equation}
  \pder{\filter{\ppvec{U}}}{t} + \ppvec{\discop}(\filter{\ppvec{U}})
  =
  \underbrace{\left(\ppvec{\discop}(\filter{\ppvec{U}}) - \filter{\nabla \cdot \ppmat{F}(\ppvec{U})} \right)
  }_{\commerror(\ppvec{U};\ppvec{\discop},\filter{(\cdot)})
  },
  \label{eq:les}
\end{equation}
where $\ppvec{\discop}(\filter{\ppvec{U}})$ denotes the numerical approximation of the exact divergence of the nonlinear fluxes evaluated for the filtered solution $\filter{\ppvec{U}}$.
This yields on the right-hand side the so-called closure term $\commerror(\ppvec{U};\ppvec{\discop},\filter{(\cdot)})$, which is a function of the full solution $\ppvec{U}$ and is thus unknown in all simulations but DNS.
It is crucial to stress that the closure term generally depends on the specific filter and the discretization scheme used to derive \cref{eq:les}.
To understand this term, we follow \cite{beck2023toward} and expand the right-hand side of \cref{eq:les} as
\begin{align}
  \begin{split}
    \pder{\filter{\ppvec{U}}}{t} + \ppvec{\discop}(\filter{\ppvec{U}}) = 
      &\phantom{+}\underbrace{\left(\ppvec{\discop}(\filter{\ppvec{U}}) - \nabla \cdot \ppmat{F}(\filter{\ppvec{U}}) \right)}_{:=\,\commerror_3} \\
    &+ \underbrace{\left(\nabla \cdot \ppmat{F}(\filter{\ppvec{U}}) - \nabla \cdot \filter{\ppmat{F}(\ppvec{U})} \right)}_{:=\,\commerror_2} \\
    &+ \underbrace{\left(\nabla \cdot \filter{\ppmat{F}(\ppvec{U})} - \filter{\nabla \cdot \ppmat{F}(\ppvec{U})} \right)}_{:=\,\commerror_1},
  \end{split}
  \label{eq:les_comm}
\end{align}
which gives rise to three distinct error terms that contribute to the overall closure term $\commerror(\ppvec{U};\ppvec{\discop},\filter{(\cdot)})$.
First, $\commerror_1$ describes the commutation error between the filter and the analytical derivative operator, which vanishes only if the filter is homogeneous, i.e.\ if its shape and filter width are identical at all points in the domain, which is usually not the case if walls and boundary layers are present.
The second term $\commerror_2$ describes the ``physical subgrid stresses'', i.e. the commutator error between the filter and the nonlinear fluxes of the NSE.
This is the term that is typically targeted when designing LES models and is always present for LES.
Lastly, the third term $\commerror_3$ describes the error introduced by the numerical divergence operator and thus depends on the applied discretization scheme.
For explicitly filtered LES, i.e.\ if the filter function is explicitly known and applied, the term $\commerror_3$ can be controlled by refining the grid while keeping the filter width fixed and it eventually vanishes for $h\rightarrow 0$.
For implicitly filtered LES in contrast, the filter is not known explicitly and is only imposed by the under-resolved discretization.
Hence, an implicitly filtered LES is by definition coarse and the commutation error $\commerror_3$ can be of the same order of magnitude as the physical subgrid stresses~\cite{moser2021statistical,ghosal1999mathematical} or even dominate the closure term in particular for low-order schemes~\cite{chow2003further}.
While this interaction between the filter and the discretization makes the analysis of implicitly filtered LES more intricate than for its explicit counterpart, it is the most commonly employed LES methodology in practice due to its computational efficiency and ease-of-use in practical applications.
Thus, we will focus on implicitly filtered LES in the following.

For LES modeling, the closure term is typically approximated by a parametrized model with parameters $\ppvec{\theta}$ that should recover the commutation error solely based on the filtered solution, i.e.\  $\model(\filter{\ppvec{U}};\ppvec{\theta}) \approx \commerror(\ppvec{U};\ppvec{\discop},\filter{(\cdot)})$.
If ML is used to learn this mapping, it is highly desirable to incorporate the fundamental symmetries of the target term into the model architecture to ensure that the model is consistent with the underlying physics~\cite{sanderse2024scientific}.
However, the interaction between the NSE, the filter and the discrete operator render it rather challenging to identify the exact form of the closure term and the symmetries it should obey as outlined in the following.

\subsection{Symmetries in Large Eddy Simulation}\label{sec:symmetries}

The continuous NSE in \cref{eq:nse} are known to exhibit a range of symmetries that can be summarized and analyzed using the framework of Lie groups~\cite{klingenberg2020symmetries}.
Symmetries are closely linked to conservation laws via Noether's theorem and encode additional structure in the solutions of PDEs.
Self-similar solutions of the NSE and scaling laws can oftentimes be derived solely from the symmetries of the governing equations~\cite{klingenberg2020symmetries}.
Among others, the NSE are known to exhibit rotational, reflectional, translational, and Galilean symmetry.
These symmetries generally only hold for the continuous NSE, but also transfer to DNS in the limit $h\rightarrow 0$, where discretization effects should become negligible for any consistent scheme.
One example of this is provided by \citet{bernardini2013turbulent}, who showed that second-order finite difference methods do not obey Galilean invariance on a discrete level, but that this error vanishes in the limit $h\rightarrow 0$.

For reduced-order simulations such as Reynolds-Averaged Navier-Stokes (RANS) or LES, knowledge of the symmetries of the filtered equations is crucial, since the applied turbulence models should retain these symmetries~\cite{ghosal1999mathematical,oberlack1997invariant}.
For RANS simulations the filter is linear, unique and depends solely on time.
This allows to derive the symmetries of the RANS equations--and consequently the symmetries the turbulence models should recover--in a straight-forward manner~\cite{klingenberg2020symmetries}.

In LES in contrast, the form and width of the applied coarse-graining filter can in principle be chosen arbitrarily.
\citet{oberlack1997invariant} showed that only a narrow class of radial filter functions leads to a set of LES equations that recovers all symmetries of the NSE, while many common filters such as the spectral cut-off filter break some of them.
This becomes even more challenging for implicitly filtered LES, since it can be shown that discretization-induced filters can be incompatible with the sole notion of applying a single, three-dimensional filter kernel~\cite{lund2003use}.
Hence, neither the filter kernel, nor the exact form of the governing equations with the induced commutation errors are known for implicitly filtered LES in the general case.
Moreover, since the discretization operator itself is part of the closure term in \cref{eq:les_comm}, the closure term for implicitly filtered LES can only be expected to recover at most the reduced set of symmetries of the discrete operator.
This makes it challenging to determine which types of symmetries an LES computation or model should actually recover.

In summary, the symmetry properties of the (unknown) filter and closure term for implicitly filtered LES are generally not known.
In practice however, simulation approaches and models are generally required to fulfill a set of (discrete) symmetries instead that correspond to the symmetries of the discretization operator.
This entails in particular a discrete rotational symmetry for 90-degree rotations, reflectional symmetry, and discrete translational symmetry along homogeneous directions with equidistant grid spacings.
We will refer to these in the following as the discrete geometric symmetries of the LES equations.
The rationale of adhering to these symmetries for data-driven turbulence modeling are three-fold.
First, the choice of how the three principal Cartesian directions in the grid are labeled should not alter the obtained results.
This directly translates to the requirement that the model and the simulation in general should be equivariant with respect to 90-degree rotations and reflections.
Second, including symmetries into the design of analytical LES models has shown to also improve their overall performance in the discrete setting~\cite{oberlack1997invariant}.
Third, imposing physical constraints in the form of symmetries into ML closure models has shown to improve their accuracy, robustness and transferability in a range of studies~\cite{shankar2023importance,toshev2023e3,wang2021incorporating}, especially in the small data regime~\cite{guan2023learning}.

\begin{figure}
  \centering
  \includegraphics{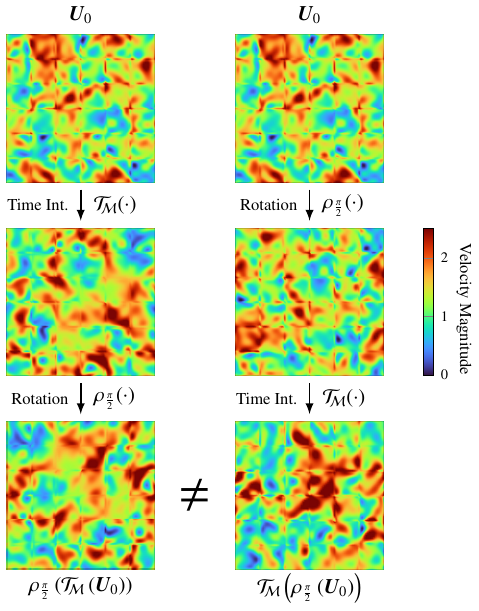}
   \caption{Missing equivariance property of a CNN-based turbulence model by showing that the operations of integrating the flow field for $t=5$ nondimensional time units including the CNN model from \cite{kurz2023deep} (denoted by $\mathcal{T}_{\!\!\!\mathcal{M}}(\cdot)$) and applying a rotation by 90 degrees $\rho_{\pi/2}(\cdot)$ do not commute, i.e. the two simulations yield different results.}\label{fig:validate_cnn_equivariance}
\end{figure}

Despite these benefits, incorporating these discrete geometric symmetries into ML-based turbulence models is still not as widespread as in traditional turbulence modeling, where their fulfillment would have been considered a hard constraint~\cite{ghosal1999mathematical,oberlack1997invariant}.
One reason for this is that the number of geometric symmetric states increases drastically from 8 to 48 when going from 2D to 3D, which makes it challenging to enforce these symmetries in a data-driven model.
This lack of symmetry-preserving properties holds in particular also for vanilla CNN-based models which are very prevalent in the literature for turbulence modeling~\cite{beck2021perspective}.
Typically, the missing symmetries of the models are then tried to be incorporated weakly using data augmentation.
For instance, the desired equivariance property for rotational symmetry would require that advancing a flow field in time and then rotating it by 90 degrees is equivalent to first rotating the flow field and then advancing it in time.
Thus, equivariance requires that the operations of integrating the flow field with the model $\mathcal{T}_{\!\!\!\model}(\ppvec{U})$ and rotating the flow field $\rho_{\pi/2}(\ppvec{U})$ commute, i.e.
\begin{equation}
  \rho_{\pi/2}\left(\mathcal{T}_{\!\!\!\model}(\ppvec{U})\right) = \mathcal{T}_{\!\!\!\model}\left(\rho_{\pi/2}(\ppvec{U})\right).
  \label{eq:equivariance}
\end{equation}
Evaluating this condition for the CNN model derived by the authors in prior work~\cite{kurz2023deep} demonstrates that \cref{eq:equivariance} does not hold, as illustrated in \cref{fig:validate_cnn_equivariance}.\footnote{The turbulent statistics computed by the LES initialized with the rotated flow field are still in good agreement with the high-fidelity simulation, but this LES yields a different instantaneous realization of the turbulent flow.}
The CNN model is thus not equivariant with respect to rotations and does not fulfill the discrete geometric symmetries of the LES equations.
We note that the same failure to adhere to the rotational invariance will occur by design for all the CNN-based closure models currently discussed in the literature, unless this property is reintroduced weakly through augmentation or soft penalties.
It is thus fair to assume that the majority of these approaches, which are highly favored for ML-driven surrogate models, fails to provide strict rotational symmetry.
The following paragraph proposes our novel modeling approach for embedding the discrete geometric symmetries into ML-based closure models using GNNs instead.
 
\section{Equivariant Turbulence Modeling}\label{sec:equivariance}
The following paragraphs introduce our key contribution for addressing these shortcomings and to provide a strategy for symmetry-preserving ML modeling using graphs and geometric deep learning.
For this, we design equivariant input and output spaces and employ GNNs to express the functional relationship between them in an equivariant manner.
First, the task of turbulence modeling is translated into a graph problem that can be addressed using equivariant GNN architectures in \cref{sec:graphs_turbmod}.
For this, a modern high-order spectral element method serves as baseline discretization, but the proposed methodology is easily applicable to other discretizations as well.
Based on this, we devise a set of suitable input and output spaces that retain these symmetries in~\cref{sec:inputs} and \cref{sec:outputs}, respectively.

\subsection{Formulating Turbulence Modeling on Graphs}\label{sec:graphs_turbmod}

\begin{figure}
  \centering
  \includegraphics{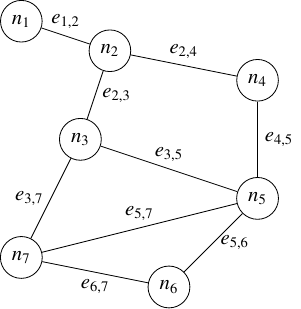}
   \caption{Exemplary outline of an undirected graph with 7 vertices $n_i\in\vertices$ and 8 edges $e_{i,j}\in\edges$. Since the graph is undirected, $e_{i,j}=e_{j,i}$ for all edges.}\label{fig:graph}
\end{figure}

In order to apply GNNs to a given problem, the task first has to be translated into an adequate graph representation.
A graph $\graph = \{\vertices, \edges\}$ describes a set of nodes $\vertices$ that are connected by a set of edges $\edges$ as illustrated in \cref{fig:graph}.
The specific structure and connectivity of the graph is encoded in its adjacency matrix $\ppmat{A}$, which is defined as
\begin{equation}
  A_{ij} = \begin{cases}
    1 & \text{if } e_{i,j}\in\edges, \\
    0 & \text{otherwise}.
  \label{eq:adjacency}
  \end{cases}
\end{equation}
Each entry $A_{ij}$ thus indicates whether the nodes $n_i, n_j\in\vertices$ are connected by an edge $e_{ij}$ or not.
In the case of an undirected graph the edges are bidirectional and the adjacency matrix becomes symmetric.

The key idea of this work is to exploit that a graph's structure is defined solely by the connectivity of its nodes and edges and does not entail information about its orientation in space unlike grid-based operators like CNNs.
For instance, applying a rotation to a graph $\graph$ as shown in \cref{fig:graph} does only rotate its visual representation in space while still retaining \emph{the same} graph $\graph$, since neither its nodes nor its edges have changed.
The same applies to any rearrangement of the nodes and edges that keeps their connectivity unchanged.
Hence, graphs are invariant under rotations, reflections and translations, which are the key geometric symmetries of the NSE as discussed in \cref{sec:symmetries}.
Hence, the central idea to this work is to apply our data-driven modeling on a graph representation to retain these symmetries by design.

Translating a discretized CFD solution into a graph is oftentimes straight-forward since the grid points and their connectivity from mesh-based simulation methods can be translated naturally into a graph representation.
Most importantly, this also holds for unstructured grids, where popular architectures like CNNs are not directly applicable~\cite{cheng2025machine}, but for which GNNs are ideally suited.
This is also a major advantage for their general applicability, but not the focus of this work.
However, scale-resolving CFD simulations can reach significant sizes with up to billions of degrees of freedom.
Training GNNs on such enormous graphs requires significant computational resources that can only be provided by distributed systems, where their efficient implementation and training become non-trivial and require careful design choices~\cite{barwey2024scalable}.
Large-scale setups as proposed for instance by~\citet{barwey2024scalable} are typically applied for cases where the GNN acts as a surrogate model for the full simulation, i.e.\ the GNN directly predicts the overall solution and thus replaces the numerical simulation altogether.
The enormous computational cost of the global GNN is justifiable for such cases, as it replaces the equally expensive numerical simulation.
However, if the GNNs are applied as turbulence model within a running CFD simulation as is the case in the present study, the considerable additional cost of global GNNs is significantly harder to justify.

\begin{figure}
  \centering
  \includegraphics{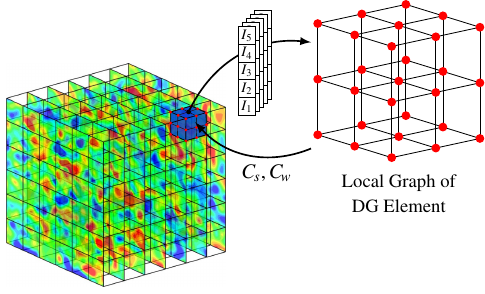}
   \caption{Outline of the local graph connectivity of a DG element in the domain. The red dots indicate the Gauss interpolation nodes, which serve as nodes in the graph representation. Edges between nodes are assumed only for direct neighbors in one of the three principal directions.}\label{fig:dg_graph}
\end{figure}

Instead, we propose to partition the global graph spanning the whole computational domain into smaller subgraphs that are then processed individually by the GNN as illustrated in \cref{fig:dg_graph}.
This allows to address the non-locality of turbulence by incorporating the vicinity of each point into the prediction, while also acknowledging that the domain of dependence at a given time instant is still confined to this vicinity due to the finite speed of sound (in compressible flow).
This approach thus represents a compromise between pointwise and global approaches.
For spectral element methods such as the DG method used in this work, each element exhibits multiple degrees of freedom, rendering it natural to define each spectral element as such a local subgraph.
Each interpolation node with the element corresponds then to a node, and edges are introduced between direct neighbors in one of the three principal directions as illustrated in \cref{fig:dg_graph}.
It is important to note that this choice of how to construct local subgraphs is arbitrary and is mostly rooted in the resulting ease of implementation for spectral element methods.
However, the general advantage of GNNs is that they are easily applicable for other discretizations, for instance with (un-)structured pointwise grids, see e.g.~\cite{dupuy2023modeling,kim2024generalizable}.
Local GNNs thus provide a very general and flexible framework for turbulence modeling that can be applied to a wide range of discretizations and problems.

\subsection{Design of Invariant Input Features}
\label{sec:inputs}

The symmetry properties of the final machine learning model depend not only on the architecture of the model, but also on the input features that are fed into the model.
To retain the discrete rotation, reflection and translation equivariance of the GNN model, its input features should be invariant under these transformations.
These restrictions are not unique to data-driven modeling, but are also considered when designing analytical turbulence models~\cite{nicoud1999subgridscale}.
For this reason, most LES models are defined with respect to the resolved velocity gradient tensor $\nabla\ppvec{u}=\partial u_i/\partial x_j$, which is a second-order tensor that can be decomposed into a symmetric and an antisymmetric component~$\nabla\ppvec{u}=\ppmat{S}+\ppmat{\Omega}$ with
\begin{equation}
  \Sij = \frac{1}{2}\left(\frac{\partial u_i}{\partial x_j} + \frac{\partial u_j}{\partial x_i}\right),
  \qquad
  \Omega_{ij} = \frac{1}{2}\left(\frac{\partial u_i}{\partial x_j} - \frac{\partial u_j}{\partial x_i}\right),
\end{equation}
representing the rate-of-strain and rate-of-rotation tensor, respectively.
A natural choice to obtain invariant input features is to use the invariants of the velocity gradient tensor as input features.
One possible set of combined invariants of the rate-of-strain and rate-of-rotation tensors was proposed by Pope~\cite{pope1975more} as
\begin{align}
  \begin{split}
    I_1 &= \text{tr}(\ppmat{S}^2),\\
    I_2 &= \text{tr}(\ppmat{\Omega}^2),\\
    I_3 &= \text{tr}(\ppmat{S}^3),\\
    I_4 &= \text{tr}(\ppmat{S}^2\ppmat{\Omega}),\\
    I_5 &= \text{tr}(\ppmat{S}^2\ppmat{\Omega}^2),
  \end{split}
  \label{eq:invariants}
\end{align}
which has been applied successfully for traditional~\cite{silvis2017physical} as well as ML-based modeling~\cite{novati2021automating,ling2016reynolds}.\footnote{Technically, the set entails a zeroth invariant $\text{tr}(\ppmat{S}) = \sum_i \frac{\partial u_i}{\partial x_i}$ that corresponds to the continuity equation (divergence-free constraint) of incompressible flow and thus vanishes in the incompressible limit. Since all flows investigated in this work are solved in the nearly incompressible regime, it is also disregarded here.}

To apply the invariants as input features for the GNN, two additional steps are required.
First, some form of nondimensionalization should be applied to the invariants to improve the transferability to physically similar problems and to ensure invariance under rescaling of the flow field.
Second, the input features should be normalized in order to improve training stability and convergence as is common practice in ML.
The velocity gradient tensor exhibits the dimension of inverse time and thus requires a reference time scale for nondimensionalization.
The nondimensionalization of the invariants in \cref{eq:invariants} then follows using the appropriate powers of the reference time scale.
A possible choice for the reference length scale can be derived from foundational quantities of turbulence--the resolved dissipation rate $\tilde{\varepsilon}$  and the effective viscosity $\nu_e=\nu+\nu_t$ encompassing the physical and turbulent viscosity, respectively.
The former can be estimated from the resolved rate-of-strain tensor as
\begin{equation}
  \tilde{\varepsilon} = 2 \nu_e \Sij \Sij .
  \label{eq:dissipation}
\end{equation}
Dimensional analysis then yields the respective time scale as\footnote{Note that $\text{tr}(\ppmat{S}\ppmat{S}) = \ppmat{S}:\ppmat{S}$ with $:$ denoting the Frobenius product, since $\ppmat{S}$ is symmetric.}
\begin{equation}
T^* = \left(\frac{\nu_e}{\tilde{\varepsilon}}\right)^{1/2}
        \overset{\eqref{eq:dissipation}}{=}
        \left(2 \Sij \Sij \right)^{-1/2}
        =
        \left(2 I_1\right)^{-1/2}.
  \label{eq:time_scale}
\end{equation}
Conveniently, multiplying the set of input features in \cref{eq:invariants} using \cref{eq:time_scale} is sufficient to scale them to an interval $[-1,1]$ for all considered applications, which is widely considered good practice for ML models.

\subsection{Design of Equivariant Predictions}\label{sec:outputs}

In a last step, the predictions of the GNN model have to be incorporated into the overall LES in a way to retain the desired equivariant properties of underlying geometric symmetries.
One established approach for this is to predict coefficients of an equivariant basis to ensure that the resulting model is equivariant under the desired transformations for all possible predicted values of the ML model~\cite{ling2016reynolds,prakash2022invariant}.
In this work, we employ a similar approach, where we select established, equivariant eddy-viscosity models as modeling basis and adapt the coefficients of these models dynamically in space and time through the GNN model to adapt it to specific flow conditions.
This approach proved to be a viable approach in previous works~\cite{kurz2023deep,novati2021automating} and can be interpreted as a special case of a more general tensor basis approach as for instance introduced in ~\cite{ling2016reynolds,prakash2022invariant}.

For eddy-viscosity models it is assumed that the subgrid-scale stress tensor $\tau_{SGS}$ is proportional to the local shear stress, i.e.
\begin{equation}
  \tau_{SGS} = -2 \nu_t \left(\ppmat{S} - \frac{1}{3}\mathrm{tr}(\ppmat{S})\, \ppmat{I}\right).
\end{equation}
Thus, this approach allows to exploit that the eddy-viscosity $\nu_t$ is a scalar and thus invariant under rotations and reflections.
In the following, we investigate two different model bases to compute the eddy viscosity in an equivariant manner, the static Smagorinsky model~(SSM)~\cite{smagorinsky1963general} and the WALE model~\cite{nicoud1999subgridscale}.
The main advantage of the WALE model is that it is designed to show the correct near-wall behavior and is thus applicable for wall-bounded flow.
Smagorinsky's model computes the eddy-viscosity as
\begin{equation}
  \nu_t = \left( \cs \, \Delta\right)^2 \sqrt{2\,\Sij\,\Sij},
  \label{eq:smago}
\end{equation}
while the WALE model computes the eddy-viscosity as
\begin{equation}
  \nu_t = \left( \cw \Delta \right)^2 \frac{ \left(\mathcal{S}^d_{ij} \mathcal{S}^d_{ij}\right)^{3/2} }{ \left(\Sij \Sij\right)^{5/2} \left(\mathcal{S}^d_{ij}\mathcal{S}^d_{ij}\right)^{5/4}},
  \label{eq:wale}\end{equation}
with
\begin{equation}
  \mathcal{S}^d_{ij} = \frac{1}{2} \left( g^2_{ij} + g^2_{ji} \right) - \frac{1}{3} \delta_{ij} g^2_{kk},
\end{equation}
based on the square of the gradient tensor $g_{ij}^2 = g_{ik} g_{kj}$.
For both models, $\Delta$ denotes the filter width, which is typically defined as
\begin{equation}
  \Delta = \frac{|V|^{\frac{1}{3}}}{N+1},
\end{equation}
where $N+1$ is the number of solution points in each spatial direction within a single DG element as also shown in \cref{fig:dg_graph}.
This normalization factor is a convention for spectral element methods to make the magnitude of $\Delta$ and thus of the $\cs$ and $\cw$ parameters in \cref{eq:smago,eq:wale} comparable to standard finite volume and finite difference methods for variable $N$.

In the following, a GNN model is trained to predict the coefficient of either the SSM following
\begin{equation}
  \cs=\model_s(I_i;\ppvec{\theta}_s,\graph),
\end{equation}
or the WALE model as
\begin{equation}
  \cw=\model_w(I_i;\ppvec{\theta}_w,\graph),
\end{equation}
based on the local invariant input features with $\ppvec{\theta}_s$ and $\ppvec{\theta}_w$ denoting the learned parameters of the GNN.
The careful design of the modeling strategy covering the input features, the model architecture and the model predictions provides a modeling framework that retains the discrete rotational, reflectional and translational symmetries of the LES equations exactly by design.
Based on this framework, the following section provides a detailed discussion of the model architecture and the design of the training loop using RL.
 
\section{Implementation Details}\label{sec:implementation}
The following paragraphs provide more details on how we implement the proposed GNN-based turbulence modeling strategy.
We first introduce Graph Convolutional Neural Networks (GCNNs) in \cref{sec:gnns} as the architecture of choice in this work.
With the architecture in place, the training setup is detailed in \cref{sec:rl:intro} and \cref{sec:rl:implementation} provides details on the implementation of the training setup.

\subsection{Graph Neural Networks}\label{sec:gnns}
Graph Neural Networks (GNNs) are a class of neural networks that operate on graphs.
While a myriad of different flavors and variants are proposed in the literature, see for instance~\cite{zhou2020graph}, they build on the same common building blocks that are introduced in the following.

\subsubsection*{Encoding initial embedding}
First, an initial embedding is computed for each node.
Given the vector of input features $\ppvec{x}_i$ for each node $n_i\in\vertices$, the embedding $\ppvec{h}_i^{(0)}$ is computed as
\begin{equation}
  \ppvec{h}_i^{(0)} = \text{Encoder}_{W}(\ppvec{x}_i), \qquad \text{for } n_i\in\vertices.
  \label{eq:encoder}
\end{equation}
Here, $\text{Encoder}_{W}(\cdot)$ denotes a (typically fully-connected) neural network that is shared across all nodes and is parameterized by the weights $\ppmat{W}$.
This encoder is used to map the input features into a latent representation at each individual node of the graph.
In a next step, message passing is performed in this latent space along the graph's edges.

\subsubsection*{Message Passing}
Most graph neural networks architectures can be expressed as a series of message passing steps.
In message passing, adjacent nodes exchange information along edges, which is then used to update the embeddings of the nodes.
Since only direct neighbors are considered in each layer of message passing, the sphere of influence for each node is limited by the number of message passing steps.

A widely applied class of GNNs are Graph Convolutional Neural Networks~(GCNNs), which were originally proposed by \citet{kipf2017semisupervised}.
GCNNs generalize the convolutional operations known from CNNs to graph-structured data by approximating their discrete convolution by a suitable message passing operation.
The message $\ppvec{m}_{i \to j}$ between the nodes $n_i$ to $n_j$ is computed in each layer $l$ as
\begin{equation}
  \ppvec{m}_{i \to j}^{(l)} = \frac{1}{\sqrt{\deg(n_i)}\sqrt{\deg(n_j)}} \ppmat{W}^{(l)} \ppvec{h}_i^{(l)},
\end{equation}
where $\deg(n_i)=\sum_{j}A_{ij} $ denotes the degree of node $n_i$, i.e.\ the number of edges connected to it, and $\ppmat{W}^{(l)}$ is the trainable weight matrix of that message passing layer.
The normalization factor is introduced to counteract the effect that the information of nodes with more connections tends to spread more prominently on the graph.
The updated embedding of a node $n_j$ is then obtained by aggregating the messages from the set of all its neighbors $\mathcal{N}(n_j)$, adding the contribution of its own embedding and applying an activation function $\sigma(\cdot)$, which yields
\begin{equation}
  \ppvec{h}_j^{(l+1)} = \sigma\left(\frac{1}{\deg(n_j)}\ppmat{W}^{(l)} \ppvec{h}_j^{(l)} + \sum_{n_i \in \mathcal{N}(n_j)} \ppvec{m}_{i \to j}^{(l)} \right).
\end{equation}
Here, the first term on the right-hand side can be interpreted as the self-connection of the node to itself $\ppvec{m}_{j \to j}^{(l)}$ and can be integrated into the summation by adding the node itself to the set of neighbors.
The summation of the messages ensures that the information from all neighbors is aggregated in a permutation-invariant way, i.e.\ the ordering of the nodes and edges does not influence the result.
It is crucial to note that the trainable weight matrix $\ppmat{W}^{(l)}$ is shared across all nodes.

Using the adjacency matrix $\ppmat{A}$ introduced in \cref{eq:adjacency}, the message passing operation can be written in a more compact form as
\begin{equation}
  \ppmat{H}^{(l+1)} = \sigma\left(\tilde{\ppmat{D}}^{-\frac{1}{2}}\tilde{\ppmat{A}}\tilde{\ppmat{D}}^{-\frac{1}{2}} \ppmat{H}^{(l)} \ppmat{W}^{(l)}\right).
  \label{eq:gcn}
\end{equation}
Here, $\ppmat{H}^{(l)}=(\ppvec{h}^{(l)}_1,\ppvec{h}^{(l)}_2,\ldots,\ppvec{h}^{(l)}_n)^T$ denotes the matrix containing the embeddings at layer $l$ of each node as rows, $\ppmat{W}^{(l)}$ is the trainable weight matrix for the layer $l$ that is shared across nodes, $\tilde{\ppmat{A}} = \ppmat{A} + \ppmat{I}$ is the adjacency matrix with added self-connections, and $\tilde{\ppmat{D}} = \mathrm{diag}(\sum_j \tilde{A}_{ij})$ denotes its diagonal degree matrix. Hence, the updated embeddings rely solely on the current embeddings on the nodes, the trainable weight matrix and the graph's structure, which is encoded in the geometry term $\tilde{\ppmat{D}}^{-\frac{1}{2}}\tilde{\ppmat{A}}\tilde{\ppmat{D}}^{-\frac{1}{2}}$.
The final application of the activation function $\sigma(\cdot)$ again should be interpreted as a element-wise operation.

\subsubsection*{Pooling and Readout}
After performing several rounds of message passing, the embeddings at all nodes have to be aggregated to extract global information from the graph, while retaining the graph's invariance under permutations of the nodes.
This means that the result of the aggregation operations must yield the same final result independent of how the nodes are ordered in their matrix representation. 
Common representatives are the average, mean, or max pooling operations, since they are all invariant under permutations.

Similar to CNNs, graph pooling operations can also be used to successively reduce the dimensionality and coarsen the graph.
Due to the inherent graph structure however, the pooling operations are less straightforward and a myriad of different approaches is presented in the literature.
Since graph pooling is not used in the following, we omit it here.

\subsubsection*{Decoder}
After the embeddings are aggregated, the final prediction of the GNN is provided by a decoder that translates the pooled embedding from the latent space into the required output dimension.
For this, several fully-connected layers are used, similarly to the initial encoding step in \eqref{eq:encoder}.

Obviously, the exact outline and ordering of these operations heavily depend on the specific problem, i.e.\ if predictions should be made for each node or for the whole graph, or even a mixture of both.
 The GNN-based turbulence model is trained in this work using RL in order to avoid model-data inconsistencies~\cite{kurz2021investigating}, which can occur when training a model \textit{offline} on a dataset without accounting for discretization effects, error accumulation, turbulent dynamics and the systematic mismatch between filtered training and actual LES data in implicit LES.
For this, the design of the RL training is given in \cref{sec:rl:intro} and \cref{sec:rl:implementation} provides more details its implementation.

\subsection{Training with Reinforcement Learning}\label{sec:rl:intro}

\begin{figure}
  \centering
  \includegraphics{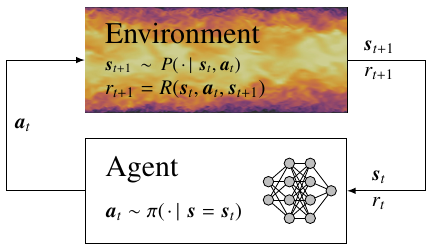} 
   \caption{Schematic of the Markov Decision Process (MDP).}\label{fig:mdp}
\end{figure}

Reinforcement Learning (RL) is a machine learning paradigm in which an agent learns how to interact with a dynamical environment in order to maximize a scalar reward signal.
For this, the concrete task is abstracted into the theoretical framework of Markov Decision Processes (MDP) illustrated in~\cref{fig:mdp}.
An MDP entails two main entities, the agent and the environment.
At each time instant $t$, the environment is in a state $\ppvec{\state}_t\in\states$, where $\states$ denotes the set of admissible environment states.
The agent observes the current state $\ppvec{\state}_t$ and based on this selects an action $\ppvec{\action}_t\in\actions$ from the set of admissible actions in this state $\actions(\ppvec{\state}_t)$.
Which action the agent takes is prescribed by the agent's policy $\pi(\ppvec{\action}_t\,|\,\ppvec{\state}_t)$, which is a probability distribution that assigns a probability to each admissible action given the current state.
The action is then sampled from the policy either randomly or greedily, depending on the specific RL algorithm.
By performing the action $\ppvec{\action}_t$, the environment transitions to a new state $\ppvec{\state}_{t+1}$ following its transition probability $P(\ppvec{\state}_{t+1}\,|\,\ppvec{\state}_t,\ppvec{\action}_t)$.
This transition probability thus encodes the dynamics of the environment.
Alongside this state transition, the agent receives a scalar reward $\reward_{t+1}$, which quantifies how good the transition was.
The reward follows from the reward function $\reward_{t+1} = \mathcal{R}(\ppvec{\state}_t,\ppvec{\action}_t,\ppvec{\state}_{t+1})$, which is typically derived with respect to some performance metric.
This process is repeated until a terminal state $\ppvec{\state}_n$ is reached yielding a trajectory of the form $\tau = \left\{ (\ppvec{\state}_0,\ppvec{\action}_0,\reward_1), (\ppvec{\state}_1,\ppvec{\action}_1,\reward_2), \ldots, (\ppvec{\state}_n) \right\}$.

The plethora of different RL algorithms proposed in the literature differ in how such trajectories are sampled and how this experience is used to derive the gradient for optimizing the agents' policy to maximize the expected cumulative reward.
In this work, the Proximal Policy Optimization (PPO) algorithm~\cite{schulman2017proximal} is employed, which is a robust policy gradient method that has demonstrated good performance in numerous of tasks with high-dimensional action spaces.
For a more detailed discussion on RL and PPO, we refer the reader to~\cite{sutton2020reinforcement,schulman2017proximal,kurz2023deep}.

The GNN architecture used in the following for the policy is summarized in \cref{tab:gnn_architecture} and combines the different building blocks outlined above.
To stress once again, the policy acts only on local subgraphs confined to a single DG element.
In contrast, the critic network incorporates the whole graph to predict the global advantage.
For this, it first performs the same operations on the local subgraphs as the policy in \cref{tab:gnn_architecture}, but then aggregates the outputs by an additional average reduction operation across the whole graph, i.e.\ across the DG elements.
The critic network used during training thus acts on the global graph to predict the correct global advantage, the policy network is only acting on the local subgraph.
Hence, global information is only used during training, while the policy itself that is later used in actual simulations acts only on the local subgraph without requiring additional communication.
It seems important to stress that in this study, RL is only applied to train the GNNs in a consistent manner within the simulations to avoid model-data-inconsistencies and is independent of our equivariant modeling strategy.

\begin{table}
  \centering
  \caption{Network architecture for the policy.}\label{tab:gnn_architecture}
  \begin{tabular}{llll}
    \toprule
    Component       & Layer & Neurons & Output Dimension \\
    \midrule
    Input           & -     & -       & $(n_{\text{nodes}},N_{\text{features}})$ \\
    \midrule
    Encoder         & MLP   & 16      & $(n_{\text{nodes}}, 16)$ \\
                    & MLP   & 32      & $(n_{\text{nodes}}, 32)$ \\
    \midrule
    Message Passing & GCNN  & 32      & $(n_{\text{nodes}}, 32)$ \\
                    & GCNN  & 32      & $(n_{\text{nodes}}, 32)$ \\
    \midrule
    Graph Readout   & -     & -       & $(32)$ \\
    \midrule
    Decoder         & MLP   & 16      & $(16)$ \\
                    & MLP   &  2      & $ (2)$ \\
    \bottomrule
  \end{tabular}
\end{table}

\subsection{Implementation}\label{sec:rl:implementation}

The RL training was performed using the PPO algorithm~\cite{schulman2017proximal} with a clipped surrogate objective.
This algorithm collects a batch of simulation trajectories and performs multiple gradient ascent on the sampled experience.
The training loop is implemented in Relexi~\cite{kurz2022relexi}, a Python package for running RL on HPC systems.
The simulations were performed with the FLEXI code~\cite{krais2021flexi}, a high-order spectral element code for CFD simulations.
The models were trained on Snellius, the Dutch national supercomputer using 8 parallel simulations on 16 compute cores each.
The training took around 12 to 48 hours per training run for the HIT case and 2 to 4 days for the channel case, depending on the test case and the chosen hyperparameters.
All codes used are open-source and available on GitHub via the links listed in the data availability statement.
For more details on the implementation, we refer the reader to our previous work~\cite{kurz2022deep,kurz2023deep,kurz2022relexi,beck2023toward}.
 
\section{Application to Homogeneous Isotropic Turbulence}\label{sec:hit}
The HIT test case is one of the most fundamental test cases in turbulence modeling and has been studied intensively in the past, also for RL-based closure models~\cite{beck2023toward,kurz2023deep,novati2021automating}.
Hence, it provides a suitable first validation case of the proposed GNN-based modeling framework.
The simulation and training setups are introduced in \cref{sec:hit:training_setup} and the obtained results are discussed in \cref{sec:hit:results}.
First, it is verified that our symmetry-preserving GNN model matches the excellent performance of traditional CNN models developed in prior work~\cite{kurz2023deep}.
Next, it is demonstrated that the GNN closure model recovers the embedded symmetries in actual LES computation in \cref{sec:hit:verification}, which the CNN model fails to do.

\subsection{Training Setup}\label{sec:hit:training_setup}

The computational setup of the HIT test case follows closely the more detailed descriptions in \cite{beck2023toward,kurz2023deep}.
The computational domain spans a cube with side length $L=2\pi$ and periodic boundary conditions in all directions.
A DNS of this flow at a Reynolds number of $\mathrm{Re}_{\lambda}\approx 180$ with respect to the Taylor microscale is computed as baseline and ground truth to assess the accuracy of the employed LES models.
The DNS is initialized with random velocity fluctuations that follow a prescribed distribution of turblent kinetic energy using Rogallo's procedure~\cite{rogallo1981numerical}.
The flow field is then advanced in time while a nodal forcing method \cite{lundgren2003linearly,delaagedemeux2015anisotropic} is employed to counter the viscous dissipation and obtain a statistically stationary flow field.
The quasi-steady DNS solution is used to compute a set of initial LES flow fields that are used to initialize the LES simulations during training.
For this, snapshots of the DNS are projected onto the LES grid exhibiting $6$ elements per direction and a polynomial degree of $N=5$, which results in a total of 36 degrees of freedom per direction and $36^3\approx \num{46700}$ degrees of freedom in total.
A set of three initial conditions extracted from the DNS at times $\tdns\in\{4,5,6\}$ are used for training the GNN model, while a single initial LES state extracted at $\tdns=8$ is kept hidden for testing.
Since the large-eddy turnover time is about $t\approx0.7$, this ensures that the initial flow state for testing is sufficiently distinct from the ones seen during training.
All results reported in the following for the different models are obtained by restarting from this hidden initial state.

\begin{table}
  \centering
  \caption{Hyperparameters of the RL training for the HIT case.}
  \label{tab:hit:hyperparameters}
  \begin{tabular}{llr}
    \toprule
    Hyperparameter                & Symbol          & Value     \\
    \midrule
    Learning rate                 & --              & $10^{-4}$ \\
    Discount factor               & $\gamma$        & $0.995$   \\
    PPO clip parameter            & $\epsilon$      & $0.2$     \\
    Policy distribution           & --              & beta~\cite{chou2017improving} \\
    Training epochs per iteration & $\nepochs$      & $3$       \\
    \midrule
    Prediction interval           & $\dtrl$         & $0.1$     \\
    Simulation time per episode   & $\tend$         & $5$       \\
    No. of parallel environments  & $\nenvs$        & $8$       \\
    Reward scaling factor         & $\alpha$        & $0.1$     \\
    Max. wavenumber for reward    & $\kmax$         & $11$      \\
    Baseline model                & --              & Smagorinsky~\cite{smagorinsky1963general}\\
    \bottomrule
  \end{tabular}
\end{table}

The most important hyperparameters used for the RL training of the HIT flow are summarized in \cref{tab:hit:hyperparameters} and the GNN model employs the architecture given in \cref{tab:gnn_architecture}.
The reward function is defined as the mean-squared error between the LES and DNS speectra normalized by an exponential function to the interval $r_t\in[-1,1]$.
The function reads as
\begin{equation}
  R(s) = 2\exp\left(-\frac{1}{\alpha \kmax}\sum_{k=1}^{\kmax}\left(\frac{\Edns(k)-\Eles(k)}{\Edns(k)}\right)^2\right)-1,
  \label{eq:hit:reward}
\end{equation}
where $\kmax$ denotes the maximum wavenumber considered in the reward computation, $\Edns$ denotes the temporally averaged target spectrum of the DNS, $\Eles(t)$ the instantaneous spectrum of the LES and $\alpha$ is a scaling parameter, see \cref{tab:hit:hyperparameters}.

\subsection{Results}\label{sec:hit:results}

\begin{figure*}
  \centering
  \includegraphics{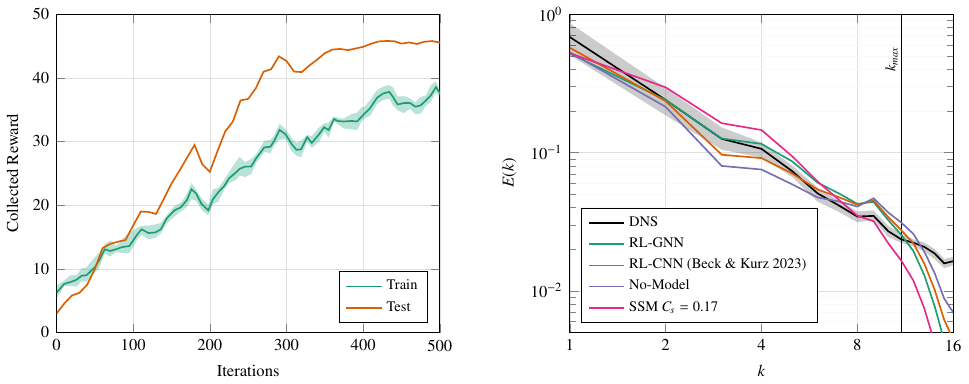}

   \caption{\textit{Left}: Evolution of collected reward during training in the training and the testing environments. The shaded area indicates the maximum and minimum reward across environments and the solid line the average. \textit{Right}: Resulting time-averaged spectra of the GNN model in $t\in[10,20]$ with the SSM with $\cs=0.17$, a no-model LES and the CNN model from \citet{beck2023toward} given for comparison. The shaded area indicates the maximum and minimum of the DNS spectrum observed during the simulation.}\label{fig:hit:reward_spectra}
\end{figure*}

The evolution of the collected reward during the training is depicted in \cref{fig:hit:reward_spectra}~(\textit{left}) showing the collected reward in the training environments as well as in the testing environment to identify overfitting during training.
The reward in the training environment increases steadily over the course of the training and converges within 500 training iterations.
The reached return is comparable to the ones observed in prior work~\cite{beck2023toward,kurz2023deep} for standard CNN architectures that also predict the $\cs$ parameter.
Interestingly, the variation between the different training environments is small and the improvement of the RL optimization very consistent indicating that the overall training setup is rather robust.

To assess the performance of the GNN model, the resulting energy spectra are compared to a no-model LES, the standard SSM and the CNN model from \cite{beck2023toward}.
The spectra are computed by running the different models until $t=20$ starting from the hidden testing state and averaging the instantaneous spectra over the time interval $t\in[10,20]$ to discard the initial transient.
Both the CNN and GNN model show superior accuracy in the medium wavenumbers in comparison to the no-model LES and the SSM.
Moreover, both show a slight increase in energy at $k\approx 9$ similar to the no-model LES to give a reasonable approximation near the last wavenumber included in the reward function ($\kmax=11$), and thus lower the overall mean squared error up to this wavenumber and maximize the collected reward.
The SSM introduces too much dissipation and shows a significant drop in energy above $k=9$.
The results in \cref{fig:hit:reward_spectra}~(\textit{right}) show that, despite small deviations, both the CNN and GNN model show excellent agreement with the DNS reference and clearly outperform the no-model LES and the SSM.

While the CNN and GNN model thus both show near perfect agreement with the DNS reference, only the novel GNN model recovers the discrete rotational and reflectional symmetry of the problem exactly as detailed in the following.

\subsection{Verification of Equivariance}\label{sec:hit:verification}

\begin{figure}
  \centering
  \includegraphics{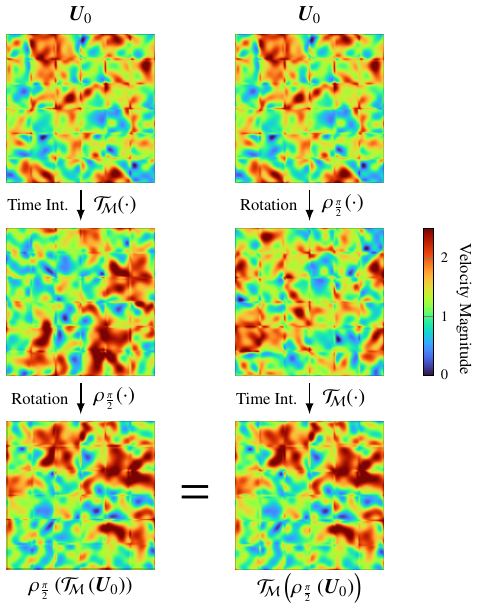}
   \caption{Demonstration of the equivariance property of the GNN-based turbulence model by showing that the operations of integrating the flow field for around 20 large-eddy turnover times (for $t=14$) including the GNN model denoted by $\mathcal{T}_{\!\!\!\mathcal{M}}(\cdot)$) and applying a rotation by 90 degrees $\rho_{\pi/2}(\cdot)$ commute, which is the definition of equivariance outlined in \cref{eq:equivariance}.}\label{fig:validate_gnn_equivariance}
\end{figure}

In a next step, it is demonstrated that the proposed GNN-based closure model does not only yield accurate results, but indeed satisfies the discrete geometric symmetries of the underlying problem, i.e.\ reflectional symmetry and discrete rotational symmetry by 90 degree increments.
For this, the same methodology as in \cref{sec:symmetries} is applied.
Two distinct LES runs are computed, both with the same initial solution $\ppvec{U}_0$ shown in \cref{fig:validate_gnn_equivariance}.
Simulation~1 is simply advanced in time to $t=14$ using the numerical scheme with the GNN-based closure model, which is denoted by $\mathcal{T}_{\!\!\!\mathcal{M}}(\ppvec{U}_0)$.
The time interval of $t\in[0,14]$ corresponds to approximately 20 large-eddy turnover times, which can be considered sufficient to demonstrate the equivariance of the GNN-based closure model.
After time-stepping, a rotation by 90 degrees is applied to the final velocity field, which is denoted by $\rho_{\pi/2}(\cdot)$.
For Simulation~2, the order of these two operations are interchanged.
First, the rotation is applied to the initial flow field and then the rotated initial condition is advanced in time using the numerical scheme with the GNN-based closure model.
As clearly visible in \cref{fig:validate_gnn_equivariance}, both operations do indeed commute as prescribed by the definition of equivariance in \cref{eq:equivariance}.
This demonstrates that the GNN-based closure model (as well as all other components of the numerical scheme) is equivariant with respect to discrete 90-degree rotations, which the CNN model was not able to satisfy as shown in \cref{fig:validate_cnn_equivariance}. 

It is important to stress that the results in \cref{fig:validate_gnn_equivariance} show minor, but visible, differences between the two velocity fields.
These do not stem from the GNN-based closure model, but are a consequence of the finite precision of floating point arithmetic and the chaotic nature of turbulence.
While both cases perform the same floating point operations they are ordered differently for the rotated and non-rotated fields, as the order is typically prescribed within the physical domain (for instance to start at the top left and move to the bottom right) and influenced by opaque compiler optimizations.
The resulting difference in rounding errors and the extreme sensitivity of turbulence to tiny perturbations causes the simulations to ultimately diverge.
It was verified that running the same test only for the solver without the GNN model yields deviations in the same order of magnitude, demonstrating that the differences do not stem from the GNN model itself.

After verifying that the equivariant GNN-model recovers the performance of CNN-based models and recovering the required discrete symmetries in practical simulations, the following section discusses the application of the GNN-based closure model to a more complex flow configuration, namely a turbulent channel flow.
 
\section{Application to Turbulent Channel Flow}\label{sec:channel}
Turbulent channel flow is a much more challenging task than the HIT case, since it exhibits boundary layer effects that have to be accounted for by the turbulence model.
First, the computational setup of the flow case and the training procedure is detailed in \cref{sec:channel:setup}.
The training behavior of the model is then discussed in \cref{sec:channel:training} before the trained models are compared against state-of-the-art analytical turbulence models in \cref{sec:channel:results}.
Finally, the learned policy of the RL-GNN model is analyzed in \cref{sec:channel:policy}.

\subsection{Training Setup}\label{sec:channel:setup}

\begin{figure}[tb]
  \centering
  \includegraphics{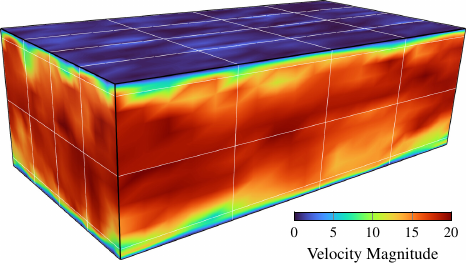}
   \caption{Instantaneous velocity magnitude in the turbulent channel flow at $\mathrm{Re}_{\tau}\approx 180$ with the white lines indicating the DG elements.}\label{fig:channel:flowfield}
\end{figure}

Turbulent channel flow is a canonical benchmark problem for wall-bounded turbulence.
It describes the flow between two parallel walls driven by a pressure gradient as shown in \cref{fig:channel:flowfield}.
This forcing can be implemented in two flavors, either by imposing a constant pressure gradient or by adapting the pressure gradient dynamically during the simulation to enforce a constant mean mass flow through the channel.
In the following, we use the former approach with a constant pressure gradient of $\mathrm{d}p/\mathrm{d}x=1$ imposed as a volume forcing term, for which the emerging mean wall shear stress follows directly as $\avg{\tau_w}=1$ via the global force balance, where we use $\avg{\cdot}$ to denote a quantity that is averaged in time and homogeneous directions in space.
The flow setup used in this work follows the setup given by~\citet{lee2015direct} at a friction Reynolds number of $\mathrm{Re}_{\tau} = 180$.
The flow is simulated in a domain of $[0, 2\pi] \times [-1, 1] \times [-\pi, \pi]$ with periodic boundary conditions in $x$- and $z$-direction and no-slip boundary conditions at the walls in $y$-direction.
The background pressure is chosen to yield a bulk Mach number of $\textrm{Ma}_b=0.3$ instead of the common value of $\textrm{Ma}_b=0.1$ to speedup the training.
For all reported LES computations, the flow is discretized using a split-form DG method with polynomial degree of $N=5$ and $4^3$ elements, which results in a rather coarse resolution of 24 solution points per spatial dimension and in total \num{13824} degrees of freedom per solution variable.
The elements are scaled in wall-normal direction such that the elements at the wall extend up to roughly $13\%$ of the channel height, while the inner elements cover the remaining $87\%$ to the channel center.
The resulting element boundary is thus located at $y^+\approx 23$.
This resolution has shown to be sufficient to yield sensible results for the baseline models and our high-order DG scheme at this Reynolds number.

\begin{table}[tb]
  \centering
  \caption{Hyperparameters of the RL training for the channel case.}
  \label{tab:channel:hyperparameters}
  \begin{tabular}{llr}
    \toprule
    Hyperparameter                & Symbol          & Value     \\
    \midrule
    Learning rate                 & --              & $10^{-4}$ \\
    Discount factor               & $\gamma$        & $0.995$   \\
    PPO clip parameter            & $\epsilon$      & $0.2$     \\
    Policy distribution           & --              & beta~\cite{chou2017improving} \\
    Training epochs per iteration & $\nepochs$      & $5$       \\
    \midrule
    Prediction interval           & $\dtrl$         & $0.2$     \\
    Simulation time per episode   & $\tend$         & $8$       \\
    No. of parallel environments  & $\nenvs$        & $8$       \\
    Reward smoothing factor       & $\rewardsmooth$ & $0.6$     \\
    Reward scaling factor         & $\alpha$        & $10^{-3}$ \\
    Baseline model                & --              & WALE~\cite{nicoud1999subgridscale}\\
    \bottomrule
  \end{tabular}
\end{table}

The RL training setup uses the same model architecture for the actor and critic networks as in the HIT case, but both receive the absolute wall distance $|d|$ as additional input feature with the invariants of the velocity gradient tensor given in~\cref{eq:invariants}.
Moreover, instead of adapting the model coefficients of the SSM, which does not vanish near walls, the RL-GNN model predicts the $\cw$ parameter of the WALE model~\cite{nicoud1999subgridscale} as given in \cref{eq:wale}.
The training is performed using $\nenvs=8$ parallel environments to sample data.
Each simulation uses a quasi-steady state of a no-model LES as initial condition and is advanced in time for $\tend=8$, which corresponds to approximately 20 flow-through times.
The model action is updated every $\dtrl=0.2$, which corresponds to about 0.5 flow-through times.
In contrast to the HIT case, turbulent channel flow is non-homogeneous and the turbulent length scales depend on the distance to the wall rendering the use of a single global energy spectrum unsuitable as target metric.
Instead, the mean velocity profile serves as the optimization target and, similar to \cref{eq:hit:reward}, the reward for the RL training is computed as
\begin{equation}
  R(s) = \exp\left(-\frac{1}{\alpha n}\sum_{i=1}^{n}\left(\frac{\Udns(y_i)-\Ules(y_i)}{\Udns(y_i)}\right)^2\right),
  \label{eq:channel:reward}
\end{equation}
where $y_i$ denotes the wall-normal coordinate of the $i$-th grid point, $n$ the total number of points in wall-normal direction up to the centerline, $\Udns(y)$ and $\Ules(y)$ are the mean wall-normal profiles of the streamwise velocity for the DNS and LES, respectively, and $\alpha$ is a scaling factor.
The mean profiles of the LES $\Ules(y)$ are obtained by time-averaging the streamwise velocity within the current actuation interval $\dtrl$ and then averaging spatially in the homogeneous $x$- and $z$-directions and the upper and lower channel halves.
Due to the short averaging time, the computed profiles are smoothed with an exponential time filter as
\begin{equation}
  \Ules = (1-\rewardsmooth)\,\Ules + \rewardsmooth\,\Unew,
\end{equation}
where $\Unew$ denotes the new profile obtained in the current actuation period and $\rewardsmooth$ denotes the smoothing factor.
This smoothing has shown to stabilize and improve the training.
On each batch of simulations, the model is trained for $\nepochs=5$ gradient steps on the whole set of trajectories, i.e.\ without mini-batching.
The remaining set of hyperparameters is summarized in \cref{tab:channel:hyperparameters}.

\subsection{Training}\label{sec:channel:training}

\begin{figure}[tb]
  \centering
  \includegraphics{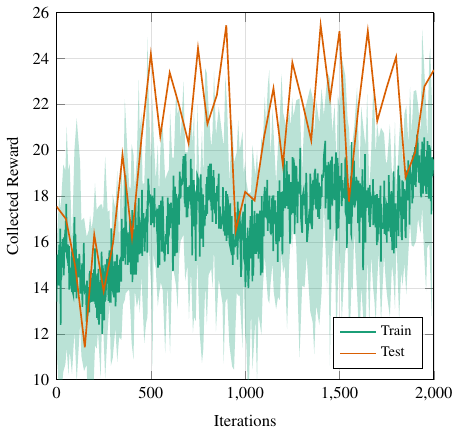}
   \caption{Evolution of collected reward during training. The shaded area indicates the maximum and minimum reward per iteration and the solid line the average over all environments. The testing run is computed with the hidden starting state and greedy evaluation of the respective policy every 50 iterations.}\label{fig:channel:training}
\end{figure}

To verify that the results are reliable, we have performed the training of the GNN model using 3 different random seeds.
While the stochastic nature of the training process leads to different training trajectories for each of them, we have verified that the final models yield similar behavior and performance for all three runs.
We demonstrate this by reporting the errors in the mean statistics for all three models in \cref{tab:channel:errors}.
However, we have selected the first seed for all the results presented in the following to keep the plots concise.

The evolution of the reward during the training of the GNN is shown in \cref{fig:channel:training}.
Clearly, the variation during the training, i.e.\ the difference between the maximum and minimum sampled reward, is considerably higher than for the HIT case.
This is to be expected since the target for the optimization for the HIT case, the energy spectrum, is a significantly more stable statistic than the mean velocity profile for the channel flow.
This means that the temporal fluctuations in the mean velocity profile are more pronounced and the profile requires significantly more samples to statistically converge.
This introduces more overall variance in the training.
Even though the reward is more volatile, \cref{fig:channel:training} clearly shows that the average training and testing rewards improve consistently during the training.

\subsection{Comparison with Analytical Models}\label{sec:channel:results}

\begin{figure*}[tb]
  \centering
  \includegraphics{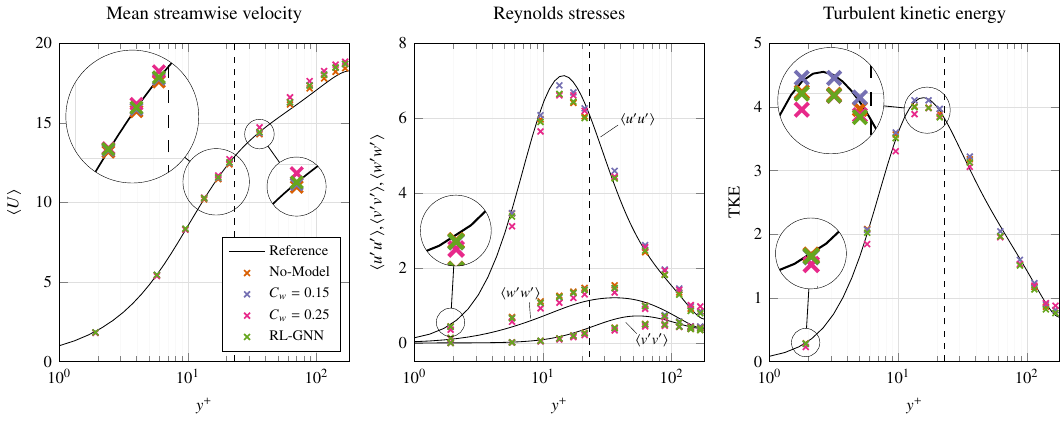}
   \caption{Averaged wall-normal profiles for the mean streamwise velocity (\textit{left}), the Reynolds stresses (\textit{center}) and the turbulent kinetic energy (\textit{right}) obtained for the RL-GNN model, a no-model LES and the standard WALE model with $\cw=0.15$ and $\cw=0.25$. Areas of special interest are highlighted and the vertical dashed line indicates the DG element boundary at $y^+\approx 23$. The errors between the models and the reference results by \citet{lee2015direct} are reported in \cref{tab:channel:errors}.}\label{fig:channel:wale}
\end{figure*}

\begin{table*}[tb]
  \centering
\sisetup{scientific-notation = true,  round-mode = figures,        round-precision = 3,         output-exponent-marker=\ensuremath{\mathrm{E}}, retain-explicit-plus = true, retain-zero-exponent = true, detect-weight,               }
  \caption{Comparison of relative mean squared errors for different models with respect to reference DNS~\cite{lee2015direct} for the channel case. In addition to the results for the distinct model also shown in \cref{fig:channel:training,fig:channel:wale,fig:channel:cs_profile}, we report the results for two more models trained from scratch using different random seeds.}\label{tab:channel:errors}
  \begin{tabular}{lccccccc}
    \toprule
    Model                      &       $\avg{u}$   &          $\avg{u'u'}$ &          $\avg{v'v'}$ &          $\avg{w'w'}$ &   TKE             \\
    \midrule
    $C_w=0.25$                 &    \num{0.014641} &        \num{0.390156} &        \num{1.741661} &        \num{0.421681} &    \num{0.130930} \\
    $C_w=0.15$                 &    \num{0.007019} &        \num{0.153266} &        \num{0.931882} &        \num{1.053540} &    \num{0.029604} \\
    No-Model                   &    \num{0.004444} &        \num{0.118650} &        \num{0.739935} &        \num{1.379753} &    \num{0.032686} \\
    \midrule
RL-GNN                     &    \num{0.006736} &        \num{0.084829} &        \num{0.964468} &        \num{1.097903} &    \num{0.032033} \\
    \midrule
    RL-GNN ($t\in[0,8]$)       &    \num{0.004920} &        \num{0.226466} &        \num{1.149642} &        \num{0.913069} &    \num{0.045581} \\
    RL-GNN (\texttt{seed=126}) &    \num{0.004818} &        \num{0.126851} &        \num{0.981748} &        \num{1.076051} &    \num{0.034482} \\
    RL-GNN (\texttt{seed=127}) &    \num{0.006623} &        \num{0.187845} &        \num{0.982686} &        \num{1.118612} &    \num{0.054299} \\
    \bottomrule
  \end{tabular}
\end{table*}

In the following we compare the results of the GNN-based turbulence model with three analytical models as baselines, which entail the WALE model with $\cw=0.25$ and $\cw=0.15$ as well as a no-model LES.
The results of the baseline models and the RL-GNN model are computed by initializing all simulations with the same a quasi-steady flow state of a no-model LES computation and advancing it in time for $t\in[0, 20]$ with the first $10$ time units discarded as warm-up time.
\Cref{fig:channel:wale} shows the mean velocity profile, the Reynolds stresses and the turbulent kinetic energy (TKE) for the different models and the reference DNS~\cite{lee2015direct}.
For a more quantitative comparison, the relative mean squared errors of the mean velocity profile, the Reynolds stresses and the TKE are reported in \cref{tab:channel:errors}.

Apparently, all models can recover the velocity profile similarly well within the viscous sublayer up to $y^+\approx 10$.
This is to be expected for two reasons.
First, the contributions of the WALE model should vanish near the wall for wall-resolved simulations and should thus become independent of $\cw$.
The second reason is that the mean velocity gradient at the wall is forced to be identical for all cases since the constant pressure gradient enforces the same mean wall-shear stress for any quasi-steady simulation.
Instead, the differences between the models become evident in the buffer and outer layer, i.e.\ for $y^+>10$.
The clear trend is that models introducing more viscosity, in particular the WALE model with $\cw=0.25$, yield a higher peak velocity and bulk velocity, while the nominally least dissipative model, the no-model LES, shows the lowest peak velocity.
The WALE model with $\cw=0.15$ falls right between the two extremes.
This is because a more viscous simulation requires a higher overall velocity difference to achieve the required wall friction of $\avg{\tau_w}=1$, resulting in a higher mass flow rate than less dissipative models.
The RL-GNN model seems to achieve an optimal balance between the two extremes.
In the buffer layer $y^+\in [10,40]$, the RL-GNN model achieves excellent agreement with the DNS reproducing the mean velocity profile almost perfectly.
At the same time, it yields a much better agreement in the outer layer than the $\cw=0.25$ model and and similar accuracy as the $\cw=0.15$ model.
Overall, the RL-GNN model shows slightly lower mean squared error for the mean velocity profile than the $\cw=0.15$ model, while being significantly more accurate than the $\cw=0.25$ model and less accurate than the no-model LES as shown in \cref{tab:channel:errors}.
However, the RL-GNN model predicts a slightly higher maximum velocity as the no-model simulation and the reference, which can be observed for all the investigated variants of the WALE model that introduce additional viscosity.
This increase in maximum velocity does only occur for the long-term evaluation of the RL-GNN.
To demonstrate this, we also report the results of the RL-GNN model for the training time interval $t\in[0,8]$ in \cref{tab:channel:errors}, where the RL-GNN model yields similar accuracy to the no-model simulations.
This might indicate that the model would benefit from a longer simulation time during training to better capture long-term effects but might also stem from the initialization of the training runs with the flow field of a no-model LES.

Next, we investigate how well the models are able to reproduce the Reynolds stresses, see \cref{fig:channel:wale} and \cref{tab:channel:errors}.
In general, the most dissipative model, i.e.\ the WALE model with $\cw=0.25$, shows a general trend to less turbulent fluctuations and TKE than the other models.
This results in the best performance for the $\cw=0.25$ model for the $\avg{w'w'}$ component, which is significantly overpredicted by all investigated models.
Other than that, the results for the $\avg{v'v'}$ and $\avg{w'w'}$ components are very similar for all models.
The largest differences between the models are visible for $\avg{u'u'}$, where again the $\cw=0.25$ model already shows clear deviations from the reference at the wall-nearest point.
The peak in the buffer layer is reproduced best by the $\cw=0.15$ model.
Interestingly, the RL-GNN model matches the DNS results the best in the outer layer $y^+>30$, where it showed clearly higher velocities in the mean velocity profile.
This might indicate that the limited resolution and degrees of freedom for the GNN requires it to do a trade-off between the accuracy of the mean velocity and the Reynolds stresses.
The same behavior can also be seen for the TKE (computed as $\mathrm{TKE} = \frac{1}{2}(\avg{u'u'} + \avg{v'v'} + \avg{w'w'})$).
The $\cw=0.25$ model shows clear deviations and underpredicts the TKE near the wall, while the $\cw=0.15$ yields the most accurate results along its peak in the buffer layer.

In summary, the RL-GNN model shows improvement over the no-model LES for the TKE and the $\avg{u'u'}$ component, while showing a more pronounced overprediction of the peak velocity at the channel center in the long-term evaluation.
The results also demonstrate that GNNs are strong candidates for subgrid scale modeling.
We have shown here that they can indeed be trained to serve as solution- and discretization-aware subgrid models while at the same time obeying the structure of the underlying formulations.
At the same time, our specific results for the channel suggest that the Reynolds number might be too low to show significant differences between the models and future work should apply this methodology to higher Reynolds numbers.
We note again that the observed behavior also shows to be qualitatively consistent for the different random seeds used for training as shown in \cref{tab:channel:errors}.

\subsection{Analysis of the Learned Policy}\label{sec:channel:policy}

\begin{figure}[tb]
  \centering
  \includegraphics{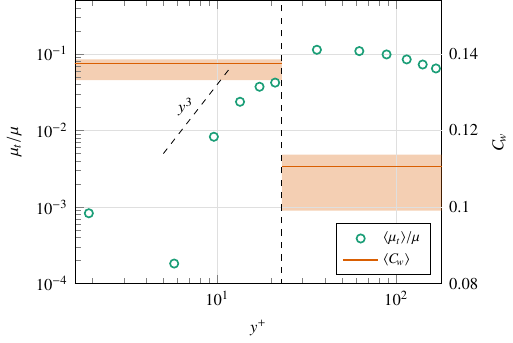}

   \caption{Mean profile of the eddy-viscosity relative to the physical viscosity $\mu_t/\mu$ and the averaged predictions for $\cw$ in the interval $t\in[10,20]$ for the RL-GNN model. The shaded areas indicate the minimum and maximum values of $\cw$ observed in that timeframe and the dashed line indicates the element boundary.}
  \label{fig:channel:cs_profile}
\end{figure}

To further investigate the characteristics of the learned policy of the RL-GNN model, \cref{fig:channel:cs_profile} shows the time-averaged eddy-viscosity profiles in wall-normal direction for the trained RL-GNN model and the corresponding predictions for the element-wise $\cw$.
Most prominently, the GNN predicts $\cw\approx 0.14$ in the elements near the wall, which cover the buffer and viscous sublayer, which is significantly higher than the predicted $\cw\approx 0.11$ for the elements in the center of the domain domain.
Moreover, the variation is significantly higher for the center elements, whereas the $\cw$ parameter is almost constant in the elements nearest at the wall.
Overall, the RL-GNN model thus learns a zonal control strategy with two distinct behaviors depending on the wall distance.

It might seem counterintuitive that $\cw$ is higher near the wall, where the flow is already highly viscous and well resolved, but it is important to note that the $\cw$ parameter is not a direct measure of the introduced viscosity, but rather a factor of proportionality that scales the eddy viscosity.
The WALE model is designed to vanish near the wall with $\mu_t \propto y^3$ scaling~\cite{nicoud1999subgridscale} for a fixed value of $\cw$.
This can also be clearly observed in \cref{fig:channel:cs_profile}, where the wall-normal profile of the predicted $\cw$ parameter and the resulting eddy viscosity $\mu_t$ are shown.
Within the viscous sublayer ($y^+<10$), $\mu_t$ becomes smaller than one per cent of the physical viscosity as the model basis of the WALE model vanishes for $y^+\rightarrow 0$.
Hence, the $\cw$ parameter can be adjusted by the model without affecting the solution quality in this region.
Within the buffer layer ($y^+>10$) however, the WALE model basis allows for non-trivial values of $\mu_t$ again.
As a consequence, the RL agent can employ the $\cw$ parameter to adjust the eddy viscosity at the 3 solution points located in the buffer layer to match the DNS results without interfering with the solution quality of the 3 solution points near the wall.
In contrast, changing $\cw$ in any of the elements in the center of the domain influences all solution points in the respective element.
This might pose too strong constraints on the model's actions to match the DNS results in the outer layer and require to predict a distribution of the $\cw$ parameter within each element as proposed in \cite{beck2023toward}.
Interestingly, the model appears to adapt the $\cw$ parameter to ensure an approximately continuous blend of the resulting mean $\mu_t$ across the element boundaries.

Overall, the RL-GNN model proves to be long-term stable and shows improvements upon the baseline reference models for some turbulent statistics of the channel flow.
This is particularly encouraging as only the mean velocity profile was used as optimization target during training, while the Reynolds stresses and the TKE were not directly optimized for.
Future work should investigate the performance of the model for higher Reynolds numbers, where the differences between the models are expected to be more pronounced.
 
\section{Conclusion}
\label{sec:conclusion}

In this work, we have proposed a novel modeling approach for symmetry-consistent ML models for LES based on GNNs.
For this, a symmetry-preserving GNN architecture and suitable input and output spaces were designed to fulfill the discrete rotational, reflectional and translational symmetries of the discretized Navier-Stokes equations.
This work demonstrated that this approach recovers these geometrical symmetries up to machine precision in actual LES.
The GNN-based modeling strategy is successfully validated to match the excellent performance of previous CNN-based models, while preserving the symmetries of the underlying physics.
To demonstrate the potential of GNNs for non-homogeneous and non-isotropic flows, the same approach was applied successfully to train a GNN-based turbulent model for turbulent channel flow.
Here, the GNN-based models learns a zonal approach with distinct behaviors in the near-wall and outer regions.
The GNN-based LES also demonstrates good agreement with DNS results for the wall-normal profiles of the mean velocity, with potential for improvement near the centerline, where the resolution is especially coarse.
More importantly, the GNN-based models also give accurate predictions of the Reynolds stresses and the turbulent kinetic energy among all models tested, while these quantities were not explicitly included as training targets.

These results show the potential of GNNs for turbulence modeling to recover physical symmetries in ML-based turbulence models without limiting their expressivity.
Our work also constitutes one of the very first applications of GNNs to the modeling of realistic, three-dimensional turbulent flows and demonstrates their trainability and usefulness in the context of in-the-loop training with modern high-order discretization schemes.
Based on these encouraging results, we see the potential for GNNs to improve the flexibility of data-driven models.
For example, the applicability of GNN-based models to unstructured grids carries the potential to improve the transferability of ML models both in terms of applicability to complex geometries and unstructured grids.
Moreover, it provides the possibility to train a single model for different discretizations and solution representations.
A drawback of GNNs that has to be addressed in future work is its increased computational cost for backpropagation, which takes around 10 times as much time per optimization step than a CNN, which might stem in parts from our custom GNN implementation.
In addition, the GNN showed a decrease in training speed within the RL context compared to CNNs.
While future research is required to address this aspect and to fully test the potential of the method, GNN architectures are an attractive machine learning model for generating mathematically consistent sub-models for the augmentation of the governing partial differential equations, taking a step towards bridging the current gap between purely data-driven and equation-agnostic approaches and classical solution schemes. 

\section*{Acknowledgments}
This work was carried out during the tenure of an ERCIM `Alain Bensoussan' Fellowship Programme.
We thank SURF (\url{www.surf.nl}) for the support in using the National Supercomputer Snellius under the NWO/EINF grant EINF-10343 for the project \texttt{relexi}.
This work is also partially supported by the project ``Discretize first, reduce next'' (with project number VI.Vidi.193.105) of Talent Programme Vidi financed by the Dutch Research Council (NWO). A.B. gratefully acknowledges funding for this work by the DFG under Germany’s Excellence Strategy EXC 2075-390740016 as well as by the European Union under the European High Performance Computing Joint Undertaking (JU) under the grant agreement No 101093393.
Moreover, the authors like to thank Syver D{\o}ving Agdestein for the insightful comments on the manuscript and the many fruitful discussions on symmetry-preserving machine learning and LES.

\section*{CRediT authorship contribution statement}
\textbf{Marius Kurz:} Conceptualization, Methodology, Software, Validation, Formal analysis, Investigation, Resources, Data curation, Writing -- original draft, Visualization, Project administration, Funding acquisition.
\textbf{Andrea Beck:} Conceptualization, Methodology, Resources, Writing -- review \& editing.
\textbf{Benjamin Sanderse:} Conceptualization, Methodology, Resources, Writing -- review \& editing, Supervision, Project administration, Funding acquisition.

\section*{Data Availability Statement}
The codes required to reproduce the results in this work are available on GitHub:
\begin{itemize}
  \item \url{https://github.com/flexi-framework/flexi} for the standard FLEXI solver (\href{https://www.gnu.org/licenses/gpl-3.0.en.html}{GPL-3.0} license)
  \item \url{https://github.com/flexi-framework/flexi-extensions/tree/smartsim} for the FLEXI version compatible with Relexi (\href{https://www.gnu.org/licenses/gpl-3.0.en.html}{GPL-3.0} license)
  \item \url{https://github.com/flexi-framework/relexi} for the Relexi framework (\href{https://www.gnu.org/licenses/gpl-3.0.en.html}{GPL-3.0} license)
  \item \url{https://github.com/m-kurz/gcnn} for the implementation of the GCNN models (\href{https://mit-license.org/}{MIT} license)
\end{itemize}
The trained models, the raw data produced in this work and detailed instructions on how to reproduce the reported results are published under the permissive \href{https://creativecommons.org/licenses/by/4.0/}{CC-BY} license at \href{https://doi.org/10.5281/zenodo.15131335}{DOI:\texttt{10.5281/zenodo.15131335}}.

\section*{Declaration of generative AI and AI-assisted technologies in the writing process.}
During the preparation of this work the authors used GitHub Copilot in order to propose linguistic improvements and \LaTeX\ typesetting in the writing process. After using this tool, the authors reviewed and edited the content as needed and take full responsibility for the content of the published article.

\bibliographystyle{elsarticle-harv}

\clearpage

\end{document}